# A Concrete Final Coalgebra Theorem for ZF Set Theory*

Lawrence C. Paulson
lcp@cl.cam.ac.uk

Computer Laboratory, University of Cambridge, England

**Abstract.** A special final coalgebra theorem, in the style of Aczel's [2], is proved within standard Zermelo-Fraenkel set theory. Aczel's Anti-Foundation Axiom is replaced by a variant definition of function that admits non-well-founded constructions. Variant ordered pairs and tuples, of possibly infinite length, are special cases of variant functions. Analogues of Aczel's Solution and Substitution Lemmas are proved in the style of Rutten and Turi [12]. The approach is less general than Aczel's, but the treatment of non-well-founded objects is simple and concrete. The final coalgebra of a functor is its greatest fixedpoint. The theory is intended for machine implementation and a simple case of it is already implemented using the theorem prover Isabelle [10].

* Thomas Forster alerted me to Quine's work. Peter Aczel and Andrew Pitts offered considerable advice and help. Daniele Turi gave advice by electronic mail. I have used Paul Taylor's macros for commuting diagrams. K. Mukai commented on the text. Research funded by the ESPRIT Basic Research Action 6453 'Types.'

# Table of Contents

# 1 Introduction

A recurring issue in theoretical computer science is the treatment of infinite computations. One important approach is based upon the final coalgebra. This category-theoretic notion relates to the methods of bisimulation and coinduction, which are heavily used in concurrency theory [6], functional programming [1] and operational semantics [7].

Aczel and Mendler [3] and also Barr [4] have proved that final coalgebras exist in set theory for large classes of naturally occurring functors. This might be supposed to satisfy most people's requirements. But Aczel [2] has argued the case for a non-standard set theory in which infinite computations, and other non-well-founded phenomena, can be modelled directly. He proposes to replace set theory's Foundation Axiom (FA) by an Anti-Foundation Axiom (AFA) that guarantees the existence of solutions to $x = \{x\}$ and more generally of all systems of equations of the form $x_i = \{x_i, x_j, \ldots\}$. His general final coalgebra theorem serves as a model construction to justify AFA.

Under AFA, a suitable functor $F$ does not merely have a final coalgebra. That final coalgebra equals $F$'s greatest fixedpoint. This is the natural dual of the theorem that a functor's initial algebra is its least fixedpoint. These fixedpoints are exact, not up to isomorphism.

The elements of the final coalgebra are easily visualised. For instance, the functor $A \times -$ (the functor $F$ such that $F(Z) = A \times Z$ on objects) yields the set of streams over $A$. The final coalgebra is also the greatest solution of $S = A \times S$. If $s \in S$ then

$$s = \langle a_1, s_1 \rangle, \quad s_1 = \langle a_2, s_2 \rangle, \quad s_2 = \langle a_3, s_3 \rangle, \ldots;$$

$s$ is the infinite stream $\langle a_1, \langle a_2, \langle a_3, \ldots \rangle\rangle\rangle$.

In standard set theory, the Foundation Axiom (FA) outlaws infinite descents under the membership relation. Under the standard definition of ordered pair we have $b \in \{a, b\} \in \langle a, b \rangle$. Infinitely nested pairs such as $s$ above would create infinite $\in$-descents, and therefore do not exist. In other words, the greatest fixedpoint of $A \times -$ is the empty set. This is not the final coalgebra (which does exist).

The approach proposed in this paper is not to change the axiom system, but instead to adopt new definitions of ordered pairs, functions, and derived concepts such as Cartesian products. Under the new definitions, the stream functor's final coalgebra is indeed its (exact) greatest fixedpoint and each stream is an infinite nest of pairs. Recursion equations are solved up to equality.

My approach handles non-well-founded tuples, and more generally ordered structures. But it does not model true non-well-founded sets, such as solutions of $x = \{x\}$. It does not work for the powerset functor, even with cardinality restrictions. I do not know whether it can express nondeterminism; one way of handling sets of outcomes may be to well-order them using the Axiom of Choice.

Aczel's book [2] puts the case for non-well-founded sets with clarity, simplicity and eloquence. Especially attractive is its presentation of four anti-foundation axioms in a uniform framework. Each axiom creates new sets and gives criteria for set equality. The axioms turn out to be pairwise incomparable; the various logicians who devised these axioms conceived four distinct notions of non-well-founded set. Is this really a fundamental notion?

I have devoted considerable effort to machine-assisted proof in ZF set theory, using the theorem prover Isabelle [8, 9]. It would be easy to separate FA from the other ZF axioms and move most of the formalisation into the resulting theory of $\mathrm{ZF}^-$. Isabelle can support parallel developments in ZF and $ZF^- + AFA$. Mechanisation of AFA requires a formalisation of the axiom and its main consequences, such as the Solution Lemma, in a form suitable for working with particular final coalgebras. A partial implementation of my approach to final coalgebras already exists [10].

*Outline.* My strategy is to construct a final coalgebra to replace AFA, and then to re-play Rutten and Turi's categorical proofs [12]. Section 2 presents basic motivation — Quine's ordered pairs and their generalisation to functions — and proves some lemmas about the cumulative hierarchy, $V_\alpha$. Section 3 defines the functor $\mathcal{Q}^I$ and its greatest fixedpoint $U^I$ and proves that $U^I$ is a final $\mathcal{Q}^I$-coalgebra. Section 4 proves the Solution and Substitution Lemmas for set equations and the special final coalgebra theorem. Section 5 discusses functors that are (or are not!) uniform on maps. Section 6 presents conclusions.

# 2 An Alternative Definition of Pairs and Functions

Let us begin with informal motivation based on the work of Quine. The following section will make formal definitions.

## 2.1 Quine's Ordered Pairs

In standard ZF set theory, the ordered pair $\langle a, b\rangle$ is defined to be $\{\{a\}, \{a, b\}\}$. The rank of $\langle a, b\rangle$ is therefore two levels above those of $a$ and $b$; there are no solutions to $b = \langle a, b\rangle$. Quine [11] has proposed a definition of ordered pair that need not entail an increase of rank. Quine's definition is complicated because (among other things) it avoids using standard ordered pairs. I regard standard pairs as indispensable, and they let us define Quine-like ordered pairs easily.

Let $\langle a, b\rangle$ denote the standard ordered pair of $a$ and $b$. Let tuples of any length consist of ordered pairs nested to the right; thus $\langle a_1, \ldots, a_n\rangle$ abbreviates $\langle a_1, \ldots, \langle a_{n-1}, a_n\rangle\rangle$ for $n > 2$. Let $A \times B$ denote the standard Cartesian product $\{\langle a, b\rangle \mid a \in A \land b \in B\}$.

Define the variant ordered pair, $\langle a; b\rangle$ by

$$\langle a; b\rangle \equiv (\{0\} \times a) \cup (\{1\} \times b). \qquad (1)$$

Note that $\langle a; b\rangle$ is just $a + b$, the disjoint sum of $a$ and $b$ (in set theory, everything is a set). The new pairing operator is obviously injective, which is a key requirement. Also, it admits non-well-founded constructions: we have $\langle 0; 0\rangle = 0$ for a start.[2]

The set equation $\langle A; z\rangle = z$ has a unique solution $z$, consisting of every (standard!) tuple of the form $\langle 1, \ldots, 1, 0, x\rangle$ for $x \in A$. The infinite stream

$$\langle A_0; A_1; \ldots; A_n; \ldots\rangle$$

is the set of all standard tuples of the form

$$\langle \underbrace{1, \ldots, 1}_{n}, 0, x\rangle$$

for $n < \omega$ and $x \in A_n$. Now $\langle a; b\rangle$ is continuous in $a$ and $b$, in the sense that it preserves arbitrary unions; thus fixedpoint methods can solve recursion equations involving variant tupling. Later we shall see that such equations possess unique fixedpoints.

Variant pairs can be generalised to a variant notion of function:

$$\tilde{\lambda}_{x\in A} b_x \equiv \bigcup_{x\in A} \{x\} \times b_x \tag{2}$$

Note that $\tilde{\lambda}_{x\in A} b_x$ is just $\Sigma_{x\in A} b_x$, the disjoint sum of a family of sets. Also note that $\langle b_0; b_1\rangle$ is the special case $\tilde{\lambda}_{i\in 2} b_i$, since $2 = \{0, 1\}$. Replacing 2 by larger ordinals such as $\omega$ gives us a means of representing infinite sequences. More generally, non-standard functions can represent infinite collections that have non-well-founded elements.

Merely replacing $\langle x, b_x\rangle$ by $\langle x; b_x\rangle$ in the usual definition of function, obtaining $\{\langle x; b_x\rangle \mid x \in A\}$, would not admit non-well-founded constructions. The rank of such a set exceeds the rank of every $b_x$. For example, if $b = \{\langle 0; b\rangle\}$ then $\{1\} \times b \in b$, violating FA; thus $b = \{\langle 0; b\rangle\}$ has no solution.

Application of variant functions is expressed using the image operator “. It is easy to check that $(\tilde{\lambda}_{x\in A} b_x) \text{``} \{a\} = b_a$ if $a \in A$. Also if $R$ is a relation with domain $A$, then $R = \tilde{\lambda}_{x\in A} R \text{``} \{x\}$; every standard relation is a variant function. The set

$$\{f \subseteq A \times \bigcup B \mid \forall_{x\in A} f \text{``} \{x\} \in B\}$$

consists of all variant functions from $A$ to $B$ and will serve as our definition of variant function space, $A \tilde{\to} B$.

Since $\tilde{\lambda}_{x\in A} b_x$ is not the function's graph, it does not determine the function's domain. For instance, $\tilde{\lambda}_{x\in A} 0 = A \times 0 = 0$. Clearly $\tilde{\lambda}_{x\in A} 0 = \tilde{\lambda}_{x\in B} 0$ for all $A$ and $B$. If $0 \in B$ then $A \tilde{\to} B$ will contain both total and partial functions: applying a variant function to an argument outside its domain yields 0.

[2] As usual in set theory, the number zero is the empty set.

## 2.2 Basic Definitions and Properties

Once we have defined the variant pairs and functions, we can substitute them in the standard definitions of Cartesian product, disjoint sum and function space. The resulting variant operators are decorated by a tilde: $\tilde{\times}$, $\tilde{+}$, $\tilde{\to}$, etc. Having both standard and variant operators is the simplest way of developing the theory. The standard operators relate the new concepts to standard set theory and they remain useful for defining well-founded constructions. But the duplication of operators may seem inelegant, and it certainly requires extra care to avoid confusing them.

**Definition 1.** The *variant ordered pair* $\langle a; b\rangle$ is defined by

$$\langle a; b\rangle \equiv (\{0\} \times a) \cup (\{1\} \times b).$$

If $\{b_x\}_{x\in A}$ is an $A$-indexed family of sets then the *variant function* $\tilde{\lambda}_{x\in A} b_x$ is defined by

$$\tilde{\lambda}_{x\in A} b_x \equiv \bigcup_{x\in A} \{x\} \times b_x$$

The *variant Cartesian product, disjoint sum* and *partial function space* between two sets $A$ and $B$ are defined by

$$\begin{aligned} A \tilde{\times} B &\equiv \{\langle x; y\rangle \mid x \in A \wedge y \in B\} \\ A \tilde{+} B &\equiv (\{0\} \tilde{\times} A) \cup (\{1\} \tilde{\times} B) \\ A \tilde{\to} B &\equiv \{f \subseteq A \times \bigcup B \mid \forall_{x\in A}\, f \text{ `` } \{x\} \in B\} \end{aligned}$$

The operators $\tilde{\times}$ and $\tilde{\to}$ can be generalised to a family of sets as usual.

**Definition 2.** If $\{B_x\}_{x\in A}$ is an $A$-indexed family of sets then their *variant sum* and *product* are defined by

$$\begin{aligned} \tilde{\sum_{x\in A}} B_x &\equiv \{\langle x; y\rangle \mid x \in A \wedge y \in B_x\} \\ \tilde{\prod_{x\in A}} B_x &\equiv \{f \subseteq A \times (\bigcup_{x\in A} \bigcup B_x) \mid \forall_{x\in A}\, f \text{ `` } \{x\} \in B_x\} \end{aligned}$$

A first attempt at exploiting these definitions is to fix an index set $I$ and solve the equation $U = I \tilde{\to} U$. There is at least one solution, namely $U = \{0\}$, since $\tilde{\lambda}_{i\in I}\, 0 = 0$. But we cannot build up variant tuples starting from 0 as we can construct the distinct sets $\{0\}$, $\{0, \{0\}\}$, .... A variant tuple whose components are all the empty set is itself the empty set.

Since $I \tilde{\to} 0 = 0$ if $I \neq 0$, one possible solution to $U = I \tilde{\to} U$ is $U = 0$. Also $I \tilde{\to} \{0\} = \{0\}$. As it happens, $U = \{0\}$ is the greatest solution.

**Proposition 3.** *If* $U = I \tilde{\to} U$ *then* $U = 0$ *or* $U = \{0\}$.

*Proof.* Suppose not, for contradiction. Then $U$ contains a non-empty element; there exist $y_0$ and $x_0$ with $y_0 \in x_0 \in U$. By the definition of $\tilde{\to}$ it follows that $y_0 = \langle i, y_1 \rangle$ where $i \in I$ and $y_1 \in x_1 \in U$ for some $x_1$. Repeating this argument yields the infinite $\in$-descent $y_0 = \langle i, y_1 \rangle$, $y_1 = \langle i, y_2 \rangle$, $y_2 = \langle i, y_3 \rangle$, …, contradicting FA. □

If tuples are to get built up, we must start with some atoms. To keep the atoms distinct from the variant tuples, each atom should contain some element that is not a (standard) pair. One atom seems sufficient. We may use 1 since by definition $1 = \{0\}$ and the empty set is not a pair. Our final coalgebra theorem will therefore be based upon the greatest solution of

$$U = \{1\} \cup (I \tilde{\to} U).$$

Some background lemmas are needed first.

### 2.3 Basic Properties of the Cumulative Hierarchy

The following results will help prove closure and uniqueness properties below.

Let $\alpha$, $\beta$ range over ordinals and $\lambda$, $\mu$ range over limit ordinals. The *cumulative hierarchy* of sets is traditionally defined by cases:

$$\begin{aligned} V_0 &= 0 \\ V_{\alpha+1} &= \mathcal{P}(V_\alpha) \\ V_\mu &= \bigcup_{\alpha<\mu} V_\alpha \end{aligned}$$

More convenient is the equivalent definition

$$V_\alpha \equiv \bigcup_{\beta<\alpha} \mathcal{P}(V_\beta). \qquad (3)$$

Kunen [5, pp. 95–7] discusses the cumulative hierarchy, using the notation $R(\alpha)$ instead of $V_\alpha$. Note some elementary consequences of these definitions:

**Lemma 4.** *If $\alpha$ is an ordinal and $\mu$ is a limit ordinal then*

$$\begin{aligned} \alpha &\subseteq V_\alpha \\ V_\alpha \times V_\alpha &\subseteq V_{\alpha+2} \\ V_\mu \times V_\mu &\subseteq V_\mu \\ V_\mu + V_\mu &\subseteq V_\mu \end{aligned}$$

It turns out that $V_\mu$ is closed under the formation of variant tuples and functions.

**Lemma 5.** *If $A \subseteq V_\mu$ and $b_x \subseteq V_\mu$ for all $x \in A$ then $\tilde{\lambda}_{x\in A} b_x \subseteq V_\mu$.*

*Proof.* This follows by the definition of $\tilde{\lambda}$, monotonicity and the facts noted above:

$$\begin{aligned}
\tilde{\lambda}_{x\in A} b_x &= \bigcup_{x\in A} \{x\} \times b_x \\
&\subseteq \bigcup_{x\in V_\mu} \{x\} \times V_\mu \\
&\subseteq V_\mu \times V_\mu \\
&\subseteq V_\mu
\end{aligned}$$

□

Thus $V_{\mu+1}$ has closure properties for variant products and sums analogous to those of $V_\mu$ for standard products and sums. It is even closed under variant function space.

**Lemma 6.** *Let $\mu$ be a limit ordinal.*

*(a) If $A \subseteq V_\mu$ then $A \tilde{\to} V_{\mu+1} \subseteq V_{\mu+1}$.*
*(b) $V_{\mu+1} \tilde{\times} V_{\mu+1} \subseteq V_{\mu+1}$.*
*(c) $V_{\mu+1} \tilde{+} V_{\mu+1} \subseteq V_{\mu+1}$.*

*Proof.* Obvious by the definitions and the previous lemma. □

These results will allow application of the Knaster-Tarski fixedpoint theorem to construct a final coalgebra. The next group of results will be used in the uniqueness proof.

**Lemma 7.** *If $A \cap V_\alpha \subseteq B$ for every ordinal $\alpha$ then $A \subseteq B$.*

*Proof.* By the Foundation Axiom, $V = \bigcup_\alpha V_\alpha$, where $V$ is the universal class. Thus $A = \bigcup_\alpha (A \cap V_\alpha)$. If $A \cap V_\alpha \subseteq B$ for all $\alpha$ then $\bigcup_\alpha (A \cap V_\alpha) \subseteq B$ and the result follows.

Using the lemma above requires some facts concerning intersection with $V_\alpha$.

**Definition 8.** A set $A$ is *transitive* if $A \subseteq \mathcal{P}(A)$.

**Lemma 9.** *$V_\alpha$ is transitive for every ordinal $\alpha$.*

*Proof.* See Kunen [5, p. 95]. □

Now we can go down the cumulative hierarchy as well as up.

**Lemma 10.** *If $\langle a, b\rangle \in V_{\alpha+1}$ then $a \in V_\alpha$ and $b \in V_\alpha$.*

*Proof.* Suppose $\langle a, b\rangle \in V_{\alpha+1}$; this is equivalent to $\{\{a\}, \{a, b\}\} \in \mathcal{P}(V_\alpha)$ and to $\{\{a\}, \{a, b\}\} \subseteq V_\alpha$. Thus $\{a, b\} \in V_\alpha$ and since $V_\alpha$ is transitive $\{a, b\} \subseteq V_\alpha$. □

**Lemma 11.** *If $\{b_x\}_{x\in A}$ is an A-indexed family of sets then*
*(a)* $(\tilde{\lambda}_{x\in A} b_x) \cap V_{\alpha+1} \subseteq \tilde{\lambda}_{x\in A}(b_x \cap V_\alpha)$
*(b)* $(\tilde{\lambda}_{x\in A} b_x) \cap V_\alpha \subseteq \bigcup_{\beta<\alpha} \tilde{\lambda}_{x\in A}(b_x \cap V_\beta)$

*Proof.* For (a) we have, by the previous lemma,

$$\begin{aligned}(\tilde{\lambda}_{x\in A} b_x) \cap V_{\alpha+1} &= \{\langle x, y\rangle \mid x \in A \wedge y \in b_x\} \cap V_{\alpha+1}\\ &\subseteq \{\langle x, y\rangle \mid x \in A \wedge y \in b_x \wedge y \in V_\alpha\}\\ &= \tilde{\lambda}_{x\in A}(b_x \cap V_\alpha).\end{aligned}$$

For (b) we have, by the definition of $V_\alpha$ and properties of unions,

$$\begin{aligned}(\tilde{\lambda}_{x\in A} b_x) \cap V_\alpha &= (\tilde{\lambda}_{x\in A} b_x) \cap \bigcup_{\beta<\alpha} \mathcal{P}(V_\beta)\\ &= \bigcup_{\beta<\alpha} (\tilde{\lambda}_{x\in A} b_x) \cap V_{\beta+1}\\ &\subseteq \bigcup_{\beta<\alpha} \tilde{\lambda}_{x\in A}(b_x \cap V_\beta).\end{aligned}$$

The last step is by (a) above. □

## 3 A Final Coalgebra

Rutten and Turi's excellent survey [12] of final semantics includes a categorical presentation of Aczel's main results. Working in the superlarge category of classes and maps between classes, they note that FA is equivalent to '$V$ is an initial $\mathcal{P}$-algebra' while AFA is equivalent to '$V$ is a final $\mathcal{P}$-coalgebra.' Put in this way, AFA certainly looks more attractive than the other anti-foundation axioms.

The present treatment of final semantics follows their development closely. Instead of assuming that $V$ is a final $\mathcal{P}$-coalgebra, we shall define a functor $\mathcal{Q}^I$, where $I$ is an arbitrary index set, and construct a final $\mathcal{Q}^I$-coalgebra, called $U^I$. The Solution and Substitution Lemmas and the Special Final Coalgebra Theorem carry over directly.

I work not in the category of classes but in the usual category **Set** of sets, which has standard functions as maps. While the former category allows certain statements to be expressed succinctly, it also requires numerous technical lemmas concerning set-based maps, etc. From the standpoint of mechanised proof, one must also bear in mind that classes have no formal existence under the ZF axioms, and class maps are two removes from existence.

### 3.1 The Functor $\mathcal{Q}$ and the Set $U$

Let $I$ be an index set, which will remain fixed throughout the paper. A typical choice for $I$ would be some limit ordinal such as $\omega$. Note that $\omega \tilde{\to} A$ contains all $\omega$-sequences over $A$; we shall find that $U^\omega$ contains all $\omega$-sequences over itself. Moreover, finite sequences can be represented by $\omega$-sequences containing infinitely many 0s, because $0 \in U^I$ (see Lemma 31 below).

**Definition 12.** The functor $\mathcal{Q}^I : \mathbf{Set} \to \mathbf{Set}$ is defined on objects by

$$\mathcal{Q}^I(A) \equiv \{1\} \cup (I \tilde{\to} A)$$

and on maps as follows. If $\pi : A \to B$ then $\mathcal{Q}^I(\pi) : \mathcal{Q}^I(A) \to \mathcal{Q}^I(B)$ satisfies

$$\mathcal{Q}^I(\pi)(1) \equiv 1$$
$$\mathcal{Q}^I(\pi)(\tilde{\lambda}_{i\in I}\, a_i) \equiv \tilde{\lambda}_{i\in I}\, \pi(a_i).$$

Reasons for this definition of $\mathcal{Q}^I$ were given after Prop. 3. It is easy to check that the functor preserves the identity map and composition. The next step is to define a set $U^I$ to be the greatest solution of $U^I = \mathcal{Q}^I(U^I)$ and prove that $U^I$ is a final $\mathcal{Q}^I$-coalgebra. Since $U^I = \{1\} \cup (I \tilde{\to} U^I)$ we may regard the elements of $U^I$ as nested $I$-indexed tuples built up from the atom 1. If some application requires a larger set of atoms, the modifications to the theory should be obvious.

To solve $U^I = \mathcal{Q}^I(U^I)$ we may apply the Knaster-Tarski fixedpoint theorem. This gives an explicit definition.

**Definition 13.** Let $\mu$ be a limit ordinal such that $I \subseteq V_\mu$. Then

$$U^I \equiv \bigcup_{Z \subseteq V_{\mu+1}} Z \subseteq \mathcal{Q}^I(Z).$$

Henceforth let us regard $I$ as fixed and drop the superscripts. The next two results indicate that $U$ really is a fixedpoint of $\mathcal{Q}$, in fact the greatest post-fixedpoint. This justifies proof by coinduction on $U$. The second result also confirms that the choice of the ordinal $\mu$ above does not matter, provided $I \subseteq V_\mu$.

**Proposition 14.** $U = \mathcal{Q}(U)$.

*Proof.* For the Knaster-Tarski theorem to apply, $\mathcal{Q}$ must be a monotone operator over the powerset of $V_\mu$. Clearly $\mathcal{Q}$ is monotone and, by Lemma 6, $\mathcal{Q}(V_{\mu+1}) \subseteq V_{\mu+1}$. □

**Proposition 15.** *If* $Z \subseteq \mathcal{Q}(Z)$ *then* $Z \subseteq U$.

*Proof.* The result follows by the definition of $U$ if we can establish $Z \subseteq V_{\mu+1}$. By Lemma 7 it suffices to prove $\forall_{z\in Z}\, z \cap V_\alpha \subseteq V_\mu$ for all $\alpha$. Proceed by transfinite induction on the ordinal $\alpha$.

Let $z \in Z$. Then $z \in \mathcal{Q}(Z) = \{1\} \cup (I \tilde{\to} Z)$. The case $z = 1$ is trivial. So we may assume $z = \tilde{\lambda}_{i\in I}\, z_i$, with $z_i \in Z$ for all $i \in I$. In this case we have

$$\begin{aligned}(\tilde{\lambda}_{i\in I}\, z_i) \cap V_\alpha &\subseteq \bigcup_{\beta<\alpha} \tilde{\lambda}_{i\in I}\, (z_i \cap V_\beta)\\ &\subseteq \bigcup_{\beta<\alpha} \tilde{\lambda}_{i\in I}\, V_\mu\\ &\subseteq V_\mu\end{aligned}$$

by Lemma 11, the induction hypothesis for $z_i$ and Lemma 5. Since $z \cap V_\alpha \subseteq V_\mu$ for all $\alpha$ we have $z \subseteq V_\mu$ for all $z \in Z$. This establishes $Z \subseteq V_{\mu+1}$. □

## 3.2 $U$ is a Final Coalgebra

To prove that $U$ is a final coalgebra requires showing that for every map $f : A \to \mathcal{Q}(A)$ there exists a unique map $\pi : A \to U$ such that $\pi = \mathcal{Q}(\pi) \circ f$:

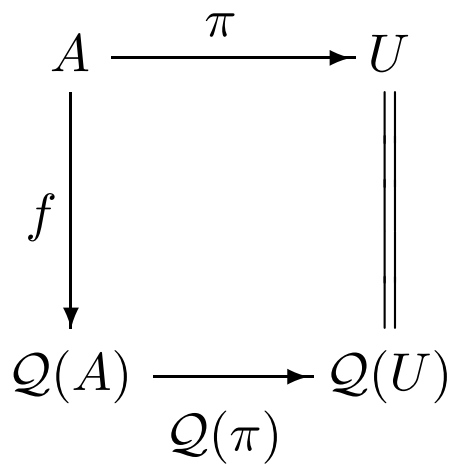

For the remainder of this section, let the set $A$ and the map $f : A \to \mathcal{Q}(A)$ be fixed.

**Lemma 16.** *There exists $\pi : A \to U$ such that $\pi(a) = \mathcal{Q}(\pi)(f(a))$ for all $a \in A$.*

*Proof.* The function $\pi$ is defined by $\pi(a) \equiv \bigcup_{n<\omega} \pi_n(a)$, where $\{\pi_n\}_{n<\omega}$ is as follows:

$$\begin{aligned} \pi_0(a) &\equiv 0 \\ \pi_{n+1}(a) &\equiv \mathcal{Q}(\pi_n)(f(a)) \end{aligned}$$

Suppose $a \in A$ and prove $\pi(a) = \mathcal{Q}(\pi)(f(a))$ by cases. If $f(a) = 1$ then the equation reduces to $1 = 1$. If $f(a) = \tilde{\lambda}_{i\in I}\, a_i$ then simple continuity reasoning establishes the equation:

$$\begin{aligned} \pi(a) &= \bigcup_{n<\omega} \pi_n(a) \\ &= \bigcup_{n<\omega} \pi_{n+1}(a) \\ &= \bigcup_{n<\omega} \tilde{\lambda}_{i\in I}\, \pi_n(a_i) \\ &= \tilde{\lambda}_{i\in I} \bigcup_{n<\omega} \pi_n(a_i) \\ &= \tilde{\lambda}_{i\in I}\, \pi(a_i) \\ &= \mathcal{Q}(\pi)(\tilde{\lambda}_{i\in I}\, a_i) \\ &= \mathcal{Q}(\pi)(f(a)) \end{aligned}$$

To show $\pi : A \to U$, use coinduction (Prop. 15). Let $Z = \{\pi(a) \mid a \in A\}$ and prove $Z \subseteq \mathcal{Q}(Z)$. If $z \in Z$ then $z = \pi(a)$ for some $a \in A$. There are two cases, as usual. If $f(a) = 1$ then $z = 1 \in \mathcal{Q}(Z)$ and if $f(a) = \tilde{\lambda}_{i\in I}\, a_i$ then $z = \tilde{\lambda}_{i\in I}\, \pi(a_i) \in \mathcal{Q}(Z)$.

Since $U$ is the greatest post-fixedpoint of $\mathcal{Q}$, this establishes $Z \subseteq U$. And since $Z$ is the range of $\pi$, this establishes $\pi : A \to U$. □

**Lemma 17.** *If* $\pi = \mathcal{Q}(\pi) \circ f$ *and* $\pi' = \mathcal{Q}(\pi') \circ f$ *then* $\pi = \pi'$.

*Proof.* Again using Lemma 7, let us use transfinite induction on the ordinal $\xi$ to prove

$$\forall_{a \in A}\, \pi(a) \cap V_\xi \subseteq \pi'(a).$$

Let $a \in A$. If $f(a) = 1$ then $\pi(a) = \pi'(a) = 1$. If $f(a) = \tilde{\lambda}_{i \in I}\, a_i$ then

$$\begin{aligned} \pi(a) \cap V_\xi &= (\tilde{\lambda}_{i \in I}\, \pi(a_i)) \cap V_\xi \\ &\subseteq \bigcup_{\eta < \xi} \tilde{\lambda}_{i \in I}\, (\pi(a_i) \cap V_\eta) \\ &\subseteq \bigcup_{\eta < \xi} \tilde{\lambda}_{i \in I}\, \pi'(a_i) \\ &= \pi'(a) \end{aligned}$$

using the hypothesis, Lemma 11, the induction hypothesis for $\eta < \xi$ and monotonicity of $\tilde{\lambda}$.

Since $\pi(a) \cap V_\xi \subseteq \pi'(a)$ for every ordinal $\xi$, we have $\pi(a) \subseteq \pi'(a)$. By symmetry we have $\pi'(a) \subseteq \pi(a)$ and therefore $\pi(a) = \pi'(a)$ for all $a \in A$. □

**Theorem 18.** *$U$ is a final $\mathcal{Q}$-coalgebra.*

*Proof.* Immediate by the previous two lemmas. □

# 4 Solutions of Equations

In his development of set theory with AFA, Aczel [2] defines systems of set-equations and proves the Solution Lemma: each system has a unique solution. Aczel introduces a class $X$ of variables and a class $V_X$ of sets built up from variables (but not themselves variables). His Substitution Lemma says that any assignment $f : X \to V$ of sets to variables can be extended to a substitution function $\hat{f} : V_x \to V$. Aczel uses these lemmas to exhibit a unique morphism for his Special Final Coalgebra Theorem.

Aczel proves the Solution and Substitution Lemmas using concrete set theory, but in Rutten and Turi's categorical presentation the proofs are much shorter. A key fact in their development is that $V$ is (assuming AFA) a final $\mathcal{P}$-coalgebra. My presentation is similar, replacing $V$ by $U$, $\mathcal{P}$ by $\mathcal{Q}$ and AFA by Theorem 18. Also I replace the category of classes by the category of sets, but this I think is only a matter of taste.

## 4.1 Preliminaries: the Binary Sum Functor

Recall that $\tilde{+}$ is the variant form of disjoint sum, defined by $A \tilde{+} B \equiv (\{0\} \tilde{\times} A) \cup (\{1\} \tilde{\times} B)$. It is a coproduct in the category **Set**, which means that for every pair of maps $f : A \to C$ and $g : B \to C$ there exists a unique map $[f, g] : A \tilde{+} B \to C$ making the diagram commute:

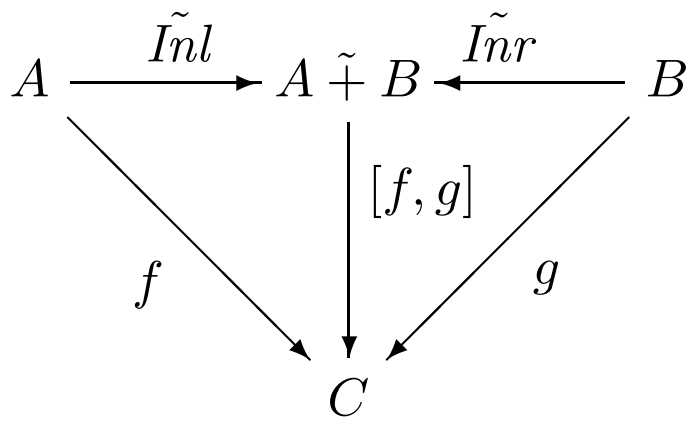

Here the injections $\widetilde{Inl} : A \to A \tilde{+} B$ and $\widetilde{Inr} : B \to A \tilde{+} B$ and the case analysis operator $[f, g]$ are defined in the obvious way.

To make $\tilde{+}$ into a functor we must define its action on maps. If $j : A \to A'$ and $k : B \to B'$ then $j \tilde{+} k : A \tilde{+} A' \to B \tilde{+} B'$ is defined (as usual) by

$$j \tilde{+} k \equiv [\widetilde{Inl} \circ j, \widetilde{Inr} \circ k]. \tag{4}$$

Some obvious properties of $[f, g]$ and of $j \tilde{+} k$ are listed below for later reference.

**Lemma 19.** *Let $f : A \to C$, $g : B \to C$, $h : C \to D$, $j : A \to A'$ and $k : B \to B'$ be maps. Then*

$$\begin{aligned}
[f, g] \circ \widetilde{Inl} &= f \\
[f, g] \circ \widetilde{Inr} &= g \\
h \circ [f, g] &= [h \circ f, h \circ g] \\
[\widetilde{Inl}, \widetilde{Inr}] &= \mathrm{id}_{A \tilde{+} B} \\
[f, g] \circ (j \tilde{+} k) &= [f \circ j, g \circ k] \\
(j \tilde{+} k) \circ \widetilde{Inl} &= \widetilde{Inl} \circ j \\
(j \tilde{+} k) \circ \widetilde{Inr} &= \widetilde{Inr} \circ k
\end{aligned}$$

### 4.2 An Expanded Version of *U*

Let $X$ be a set of variables or indeterminates for use in equations. The set $U_X^I$ is constructed in the same way as $U^I$ except that each level contains a copy of $X$. Thus an element of $U_X^I$ is just like an element of $U^I$ except that it may contain elements of $X$ at each stage in its construction. In formalizing equations between sets, each left-hand side will consist of a variable from $X$ while each right-hand side will be drawn from $U_X^I$. The definition of $U_X^I$ closely resembles that of $U^I$.

**Definition 20.** Let $\mu$ be a limit ordinal such that $X \cup I \subseteq V_\mu$. Then

$$U^I \equiv \bigcup_{Z \subseteq V_{\mu+1}} Z \subseteq \mathcal{Q}^I(X \tilde{+} Z).$$

Let us again drop the superscript $I$. The proof of the following result is omitted because of its similarity to the proof for $U$.

**Proposition 21.** *Let $U_X$ be defined as above. Then*

*(a)* $U_X = \mathcal{Q}(X \tilde{+} U_X)$.
*(b) If* $Z \subseteq \mathcal{Q}(X \tilde{+} Z)$ *then* $Z \subseteq U_X$.
*(c)* $U_X$ *is the final coalgebra for the functor* $\mathcal{Q}(X \tilde{+} -)$.

*Proof.* See Appendix A.

### 4.3 An Embedding

There is an obvious embedding $\sigma_X : U \to U_X$ that copies an element of $U$ into $U_X$ and never introduces an element of $X$:

$$\begin{aligned} \sigma_X(1) &= 1 \\ \sigma_X(\tilde{\lambda}_{i \in I}\, u_i) &= \tilde{\lambda}_{i \in I}\, \tilde{Inr}(\sigma_X(u_i)) \end{aligned}$$

The equations can be summarised neatly by $\sigma_X = \mathcal{Q}(\tilde{Inr} \circ \sigma_X)$. The embedding will be useful for creating equations with constant right-hand sides.

Although the embedding is obvious, its existence deserves to be proved. Aczel derives the analogous embedding from $V$ into $V_X$ by direct recourse to AFA [2]. Rutten and Turi [12] omit this step, which in their categorical style might be done by showing that $V_X$ is a final coalgebra for the functor $\mathcal{P}(X + -)$.

**Lemma 22.** *There exists a unique map* $\sigma_X : U \to U_X$ *such that*

$$\sigma_X = \mathcal{Q}(\tilde{Inr} \circ \sigma_X).$$

*Proof.* Recalling the equation $U_X = \mathcal{Q}(X \tilde{+} U_X)$, consider the following diagram:

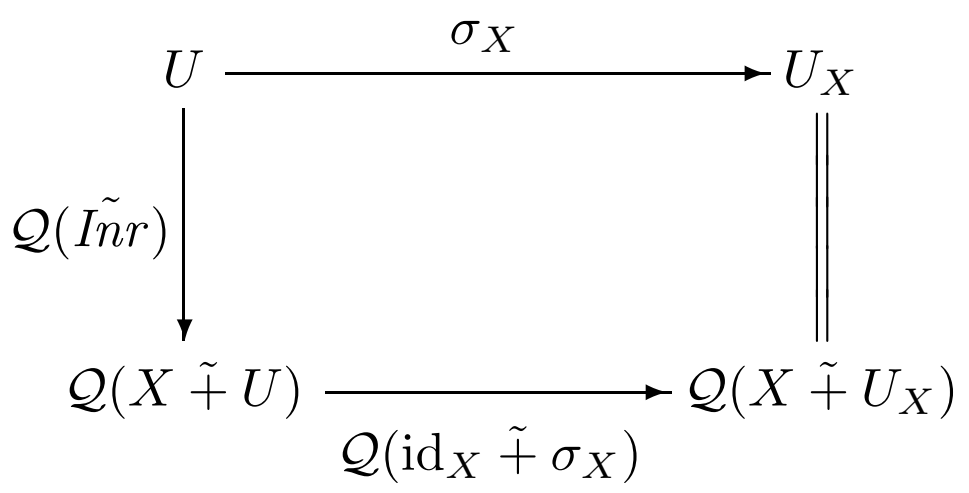

Since $(U, \mathcal{Q}(\tilde{Inr}))$ is a coalgebra for $\mathcal{Q}(X \tilde{+} -)$ and $U_X$ is a final coalgebra, there exists a unique map $\sigma_X$ such that the diagram commutes. Now

$$\begin{aligned} \sigma_X &= \mathcal{Q}(\mathrm{id}_X \tilde{+} \sigma_X) \circ \mathcal{Q}(\tilde{Inr}) \\ &= \mathcal{Q}((\mathrm{id}_X \tilde{+} \sigma_X) \circ \tilde{Inr}) \\ &= \mathcal{Q}(\tilde{Inr} \circ \sigma_X) \end{aligned}$$

by Lemma 19. □

### 4.4 Substitution

Let $f : X \to U$ be a function. Then the substitution function $\hat{f} : U_X \to U$ essentially copies its argument, replacing everything of the form $\widetilde{Inl}(x)$ by $f(x)$ for $x \in X$. This case analysis can be expressed with the help of the $[f, \hat{f}]$ notation:

$$\begin{aligned} \hat{f}(1) &= 1 \\ \hat{f}(\tilde{\lambda}_{i \in I}\, z_i) &= \tilde{\lambda}_{i \in I}\, [f, \hat{f}](z_i) \end{aligned}$$

These two equations can be expressed succinctly by $\hat{f} = \mathcal{Q}([f, \hat{f}])$.

Clearly, substitution over a 'term' containing no 'variables' can have no effect. The formal statement of this fact justifies calling $\sigma_X$ an embedding.

**Lemma 23 (Embedding).** *Let $f : X \to U$ and $g : U_X \to U$ be functions. If $g = \mathcal{Q}([f, g])$ then $g \circ \sigma_X = \mathrm{id}_U$.*

*Proof.* By Lemma 22 and Lemma 19 we have

$$\begin{aligned} g \circ \sigma_X &= \mathcal{Q}([f, g]) \circ \mathcal{Q}(\widetilde{Inr} \circ \sigma_X) \\ &= \mathcal{Q}([f, g] \circ \widetilde{Inr} \circ \sigma_X) \\ &= \mathcal{Q}(g \circ \sigma_X). \end{aligned}$$

Since $U = \mathcal{Q}(U)$, the following diagram commutes:

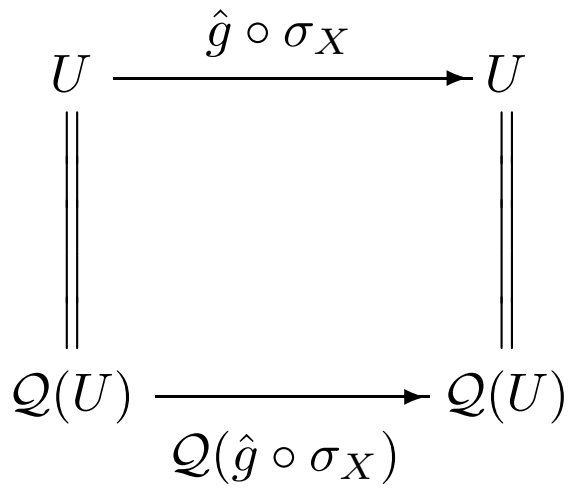

Since $U$ is the final $\mathcal{Q}$-coalgebra, it has only one homomorphism into itself, namely the identity. This yields $g \circ \sigma_X = \mathrm{id}_U$. $\square$

### 4.5 Solution and Substitution Lemmas

If $X$ is a set of variables then a function $\nu : X \to U_X$ defines a system of equations of the form $x = \nu_x$ for all $x \in X$. Such a system has a unique solution $f : X \to U$ such that $f(x) = \hat{f}(\nu_x)$ for $x \in X$. More concisely, a solution satisfies $f = \hat{f} \circ \nu$.

**Lemma 24 (Solution).** *Let $\nu : X \to U_X$ be a function. There exist unique functions $f : X \to U$ and $\hat{f} : U_X \to U$ such that*

$$f = \hat{f} \circ \nu \quad \textit{and} \quad \hat{f} = \mathcal{Q}([f, \hat{f}]).$$

*Proof.* Recalling the equation $U_X = \mathcal{Q}(X \tilde{+} U)$, consider the following diagram:

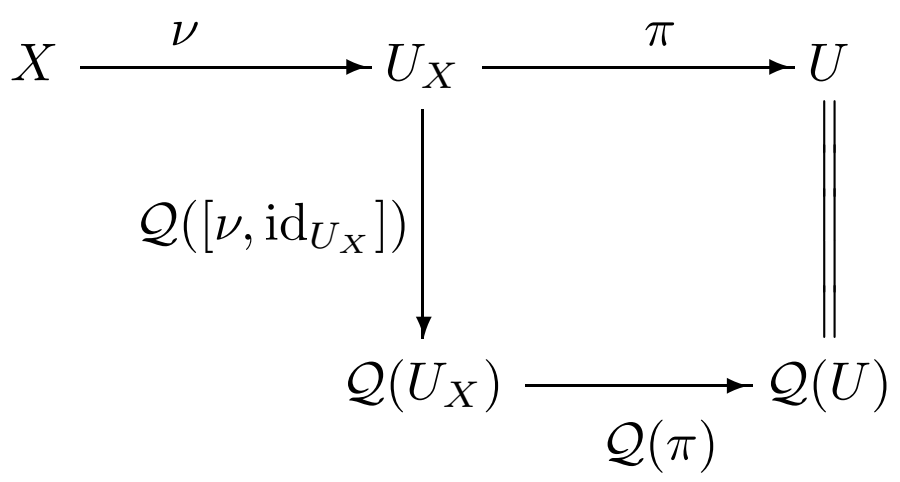

Since $(U_X, \mathcal{Q}([\nu, \mathrm{id}_{U_X}]))$ is a $\mathcal{Q}$-coalgebra and $U$ is a final $\mathcal{Q}$-coalgebra, there exists a unique map $\pi$ such that the diagram commutes. By Lemma 19 we have

$$\begin{aligned}\pi &= \mathcal{Q}(\pi) \circ \mathcal{Q}([\nu, \mathrm{id}_{U_X}]) \\ &= \mathcal{Q}(\pi \circ [\nu, \mathrm{id}_{U_X}]) \\ &= \mathcal{Q}([\pi \circ \nu, \pi \circ \mathrm{id}_{U_X}]) \\ &= \mathcal{Q}([\pi \circ \nu, \pi])\end{aligned}$$

Putting $\hat{f} = \pi$ and $f = \pi \circ \nu$ yields the desired functions. As for uniqueness, if $f = \hat{f} \circ \nu$ and $\hat{f} = \mathcal{Q}([f, \hat{f}])$ then $\hat{f} = \pi$ by finality of $U$. □

In this proof, note that $\mathcal{Q}([\nu, \mathrm{id}_{U_X}])$ substitutes using $\nu$ but only to depth one. The following lemma justifies the $\hat{f}$ notation for substitution by $f$. The idea is to convert $f : X \to U$ into a trivial system of equations, then solve them.

**Lemma 25 (Substitution).** *Let $f : X \to U$ be a function. There exists a unique function $\hat{f} : U_X \to U$ such that $\hat{f} = \mathcal{Q}([f, \hat{f}])$.*

*Proof.* Consider the composed map $\sigma_X \circ f : X \to U_X$. Apply the Solution Lemma with $\nu = \sigma_X \circ f$, obtaining maps $g : X \to U$ and $\hat{g} : U_X \to U$ such that $g = \hat{g} \circ \nu$ and $\hat{g} = \mathcal{Q}([g, \hat{g}])$. Now

$$g = \hat{g} \circ \sigma_X \circ f = f$$

by Lemma 23. Putting $\hat{f} = \hat{g}$ we obtain $\hat{f} = \mathcal{Q}([f, \hat{f}])$. Uniqueness follows by the uniqueness property of the Solution Lemma. □

I should prefer to prove the Substitution Lemma earlier, but the simplest proof seems to rely on the Solution Lemma. Turi has pointed out (by electronic mail) that the Substitution Lemma has a trivial proof if $U_X$ is defined to be an initial algebra rather than a final coalgebra. But then $U_X$ would contain only finite constructions; the embedding $\sigma_X : U \to U_X$ would not exist; non-well-founded objects obtained via the Solution Lemma could not participate in further set equations.

### 4.6 Special Final Coalgebra Theorem

The main theorem applies to functors that are uniform on maps. This notion is due to Aczel [2], but I follow Rutten and Turi's [12] formulation.

We shall no longer work in the category **Set** of sets but rather in the full subcategory $\mathbf{Set}_U$ whose objects are the subsets of $U$. Recall that $U$, in turn, depends upon the choice of index set $I$; we can make $U$ as large as necessary.

**Definition 26.** A functor $F : \mathbf{Set}_U \to \mathbf{Set}_U$ is *uniform on maps* if for all $A \subseteq U$ there exists a mapping $\phi_A : F(A) \to U_A$ such that

$$F(h) = \hat{h} \circ \phi_A \quad \text{for all } h : A \to U.$$

The mapping $\phi_A$ is called the $U_A$ *translation.*

Let us only consider functors that preserve inclusion maps. This is a natural restriction since all functors preserve identity maps, and inclusion maps are identity maps when regarded as sets. All such functors on $\mathbf{Set}_U$ have fixedpoints.

**Lemma 27.** *If the functor* $F : \mathbf{Set}_U \to \mathbf{Set}_U$ *preserves inclusions then there exists an object* $J_F : \mathbf{Set}_U$ *such that* $J_F$ *is the greatest fixedpoint and greatest post-fixedpoint of* $F$*.*

*Proof.* Apply the Knaster-Tarski fixedpoint theorem to the lattice of subsets of $U$. The functor $F$ is necessarily monotone because it preserves inclusions; if $\iota : A \to B$ then $F(\iota) : F(A) \to F(B)$; thus if $A \subseteq B$ then $F(A) \subseteq F(B)$. □

**Theorem 28 (Special Final Coalgebra).** *If the functor* $F : \mathbf{Set}_U \to \mathbf{Set}_U$ *preserves inclusions and is uniform on maps then* $J_F$ *is a final* $F$*-coalgebra.*

*Proof.* Let $(A, f)$ be an $F$-coalgebra. We must exhibit a unique map $h : A \to J_F$ such that $h = F(h) \circ f$:

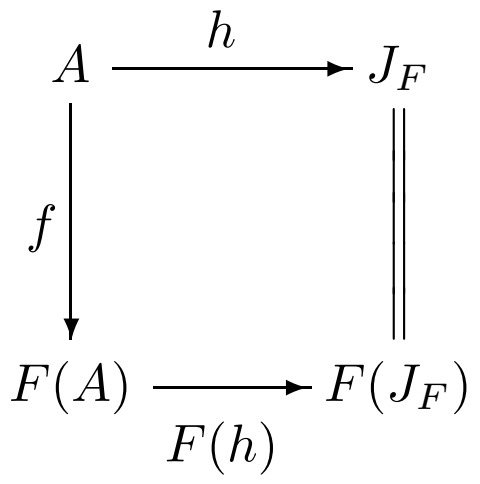

Since $F$ is uniform on maps, $h = F(h) \circ f$ is equivalent to $h = \hat{h} \circ \phi_A \circ f$. Such a map $h$ is precisely a solution of the system of equations $a = \phi_A(f(a))$ for $a \in A$. Applying the Solution Lemma with $\nu = \phi_A \circ f$, we obtain a unique map $h : A \to U$ such that $h = F(h) \circ f$.

A standard coinduction argument proves $h : A \to J_F$. Writing $h``A$ for the image of $A$ under $h$, we have

$$\begin{aligned} h``A &= (F(h) \circ f)``A \\ &= F(h)``(f``A) \\ &\subseteq F(h)``F(A) \\ &\subseteq F(h``A) \end{aligned}$$

since $h : A \to h``A$ and $F(h) : F(A) \to F(h``A)$.

The range of $h$ is thus a post-fixedpoint of $F$ and is contained in the greatest post-fixedpoint, namely $J_F$. □

# 5 Functors Uniform on Maps

If $F$ is uniform on maps then, in essence, its effect upon a map $h : A \to U$ can be expressed as the substitution of $h$ over a pattern derived from the argument; if $b \in F(A)$ then $F(h)(b) = \hat{h}(\phi_A(b))$. Most natural functors are uniform on maps but there is at least one glaring exception. Let us examine some typical cases, starting with a very easy one.

## 5.1 The Constant Functor

If $C \subseteq U$ then let $\mathrm{K}_C$ be the constant functor such that $\mathrm{K}_C(A) = C$ for all $A : \mathbf{Set}_U$ and such that $\mathrm{K}_C(f) = \mathrm{id}_C$ for all maps $f : A \to A'$.

**Proposition 29.** *If $C : \mathbf{Set}_U$ then the constant functor $\mathrm{K}_C : \mathbf{Set}_U \to \mathbf{Set}_U$ is uniform on maps.*

*Proof.* Let $A$ be a set such that $A : \mathbf{Set}_U$. Define $\phi_A : C \to U_A$ by $\phi_A(c) = \sigma_A(c)$ for all $c \in C$. Now

$$\mathrm{K}_C(h)(c) = c = (\hat{h} \circ \sigma_A)(c) = (\hat{h} \circ \phi_A)(c)$$

for all $c \in C$ by Lemma 23. □

## 5.2 Binary Product

The set $U$ satisfies the inclusion $U \tilde{\times} U \subseteq U$. So it is easy to see that $\tilde{\times} : \mathbf{Set}_U \times \mathbf{Set}_U \to \mathbf{Set}_U$ is a functor when extended to maps in the standard way. If $f : A \to A'$ and $g : B \to B'$ are maps then $f \tilde{\times} g : A \tilde{\times} B \to A' \tilde{\times} B'$ is the map that takes $\langle a; b\rangle$ to $\langle f(a); g(b)\rangle$.

**Proposition 30.** *If $F$, $G : \mathbf{Set}_U \to \mathbf{Set}_U$ are uniform on maps, then the functor*

$$F(-) \tilde{\times} G(-) : \mathbf{Set}_U \to \mathbf{Set}_U$$

*is uniform on maps.*

*Proof.* Let $A$ be a set such that $A : \mathbf{Set}_U$, or equivalently $A \subseteq U$. Clearly we have $F(A) \tilde{\times} G(A) : \mathbf{Set}_U$. Since $F$ and $G$ are uniform on maps there exist $U_A$ translations

$$\begin{aligned} &\phi_A : F(A) \to U_A \text{ such that } F(h) = \hat{h} \circ \phi_A, \quad \text{and} \\ &\psi_A : G(A) \to U_A \text{ such that } G(h) = \hat{h} \circ \psi_A \end{aligned}$$

for all $h : A \to U$. We must define a $U_A$ translation for $F(-) \tilde{\times} G(-)$.

Let $\theta_A = \phi_A \tilde{\times} \psi_A$. Thus $\theta_A(\langle b; c\rangle) = \langle \phi_A(b); \psi(b)\rangle$ for all $b \in F(A)$ and $c \in G(A)$. Now

$$\begin{aligned} (F(-) \tilde{\times} G(-))(h) &= F(h) \tilde{\times} G(h) \\ &= (\hat{h} \circ \phi_A) \tilde{\times} (\hat{h} \circ \psi_A) \\ &= (\hat{h} \tilde{\times} \hat{h}) \circ (\phi_A \tilde{\times} \psi_A) \\ &= \hat{h} \circ (\phi_A \tilde{\times} \psi_A) \\ &= \hat{h} \circ \theta_A \end{aligned}$$

and $\theta_A$ is the desired $U_A$ translation. Replacing $\hat{h} \tilde{\times} \hat{h}$ by $\hat{h}$ above is valid because $\hat{h} \tilde{\times} \hat{h}$ is applied only to variant pairs in that context. □

### 5.3 Binary Sum

The proposition about $\tilde{+}$ resembles the one about $\tilde{\times}$ presented above, but first we have to show that $U$ is closed under $\tilde{+}$.

**Lemma 31.** *$U \tilde{+} U \subseteq U$.*

*Proof.* Since $U \tilde{+} U = (\{0\} \tilde{\times} U) \cup (\{1\} \tilde{\times} U)$ and $U$ is closed under $\tilde{\times}$, it suffices to show $\{0, 1\} \subseteq U$. By coinduction (Prop. 15) it suffices to show $\{0, 1\} \subseteq \mathcal{Q}(\{0, 1\})$. This holds because

$$\mathcal{Q}(\{0, 1\}) = \{1\} \cup (I \tilde{\to} \{0, 1\})$$

and $0 = \tilde{\lambda}_{i \in I}\, 0 \in I \tilde{\to} \{0, 1\}$. □

Recall that $\tilde{+}$ is a functor on **Set** whose effect on maps was described in Sect. 4.1. Now we know that $\tilde{+}$ is also a functor in the full subcategory $\mathbf{Set}_U$.

**Proposition 32.** *If $F$, $G : \mathbf{Set}_U \to \mathbf{Set}_U$ are uniform on maps, then the functor*

$$F(-) \tilde{+} G(-) : \mathbf{Set}_U \to \mathbf{Set}_U$$

*is uniform on maps.*

*Proof.* Let $A$ be a set such that $A : \mathbf{Set}_U$. Then $F(A) \tilde{+} G(A) : \mathbf{Set}_U$ and there exist $U_A$ translations

$$\phi_A : F(A) \to U_A \text{ such that } F(h) = \hat{h} \circ \phi_A, \quad \text{and}$$
$$\psi_A : G(A) \to U_A \text{ such that } G(h) = \hat{h} \circ \psi_A$$

for all $h : A \to U$.

Let $\theta_A = \phi_A \tilde{+} \psi_A$. Then

$$\begin{aligned}(F(-) \tilde{+} G(-))(h) &= F(h) \tilde{+} G(h) \\ &= (\hat{h} \circ \phi_A) \tilde{+} (\hat{h} \circ \psi_A) \\ &= (\hat{h} \tilde{+} \hat{h}) \circ (\phi_A \tilde{+} \psi_A) \\ &= \hat{h} \circ (\phi_A \tilde{+} \psi_A) \\ &= \hat{h} \circ \theta_A\end{aligned}$$

and $\theta_A$ is the desired $U_A$ translation. □

### 5.4 Sum of a Family of Sets

Let $\{B_x\}_{x \in C}$ be a $C$-indexed family of sets. If $C \subseteq U$ and $B_x \subseteq U$ for all $x \in C$ then we have $\tilde{\sum}_{x \in C} B_x \subseteq U$. Note that $\tilde{\sum}_{x \in C} B_x$ is the usual generalisation of $C \tilde{\times} B$ to allow $B$ to depend upon $x \in C$; the two functors have the same effect upon maps. But the proposition about $\tilde{\sum}$ differs in one key respect from that about $\tilde{\times}$: the index set is not given by a functor but is constant.

**Proposition 33.** *If $C : \mathbf{Set}_U$ and $\{F_x : \mathbf{Set}_U \to \mathbf{Set}_U\}_{x \in C}$ is a $C$-indexed family of functors that are uniform on maps, then the functor*

$$\tilde{\sum_{x \in C}} F_x(-) : \mathbf{Set}_U \to \mathbf{Set}_U$$

*is uniform on maps.*

*Proof.* Let $A$ be a set such that $A : \mathbf{Set}_U$. Then $\tilde{\sum}_{x \in C} F_x(A) : \mathbf{Set}_U$. There exists a $U_A$ translation $\phi_{x,A} : F_x(A) \to U_A$ such that $F_x(h) = \hat{h} \circ \phi_{x,A}$ for all $x \in A$ and $h : A \to U$.

The $U_A$ translation for $\tilde{\sum}_{x \in C} F_x(-)$, called $\theta_A$, is defined by

$$\theta_A(\langle x; y \rangle) = \langle \sigma_A(x), \phi_{x,A}(y) \rangle$$

for all $x \in C$ and $c \in F_x(A)$. Recall that $\sigma_A$ is the inclusion map from $U$ into $U_A$. Now by Lemma 23 we have

$$\begin{aligned}\hat{h}(\theta_A(\langle x; y \rangle)) &= \hat{h}(\langle \sigma_A(x), \phi_{x,A}(y) \rangle) \\ &= \langle \hat{h}(\sigma_A(x)), \hat{h}(\phi_{x,A}(y)) \rangle \\ &= \langle x, F_x(h) \rangle \\ &= (\tilde{\sum_{x \in C}} F_x(h))(\langle x; y \rangle)\end{aligned}$$

which proves $\hat{h} \circ \theta_A = (\tilde{\sum\limits_{x \in C}} F_x(-))(h)$. □

## 5.5 Product of a Family of Sets

Again let $\{B_x\}_{x \in C}$ be a $C$-indexed family of sets. If $C \subseteq I$ (not $C \subseteq U$ as above!) and $B_x \subseteq U$ for all $x \in C$ then $\tilde{\prod\limits_{x \in C}} B_x \subseteq I \tilde{\to} U \subseteq U$.

Thus $\tilde{\prod} : \mathbf{Set}_U^I \to \mathbf{Set}_U$ is a functor whose effect on maps is as follows. If $\{f_x : B_x \to B'_x\}_{x \in C}$ is a $C$-indexed family of maps then

$$\tilde{\prod_{x \in C}} f_x : \tilde{\prod_{x \in C}} B_x \to \tilde{\prod_{x \in C}} B'_x$$

is the usual pointwise map that takes $\tilde{\lambda}_{x \in C} b_x$ to $\tilde{\lambda}_{x \in C} f_x(b_x)$.

**Proposition 34.** *If $C \subseteq I$ and $\{F_x : \mathbf{Set}_U \to \mathbf{Set}_U\}_{x \in C}$ is a $C$-indexed family of functors that are uniform on maps, then the functor*

$$\tilde{\prod_{x \in C}} F_x(-) : \mathbf{Set}_U \to \mathbf{Set}_U$$

*is uniform on maps.*

*Proof.* Let $A$ be a set such that $A : \mathbf{Set}_U$. Then $\tilde{\prod\limits_{x \in C}} F_x(A) : \mathbf{Set}_U$. There exists a $U_A$ translation $\phi_{x,A} : F_x(A) \to U_A$ such that $F_x(h) = \hat{h} \circ \phi_{x,A}$ for all $x \in A$ and $h : A \to U$.

Let $\theta_A = \tilde{\prod\limits_{x \in C}} \phi_{x,A}$. Now

$$\begin{aligned}
(\tilde{\prod_{x \in C}} F_x(-))(h) &= \tilde{\prod_{x \in C}} F_x(h) \\
&= \tilde{\prod_{x \in C}} \hat{h} \circ \phi_{x,A} \\
&= (\tilde{\prod_{x \in C}} \hat{h}) \circ (\tilde{\prod_{x \in C}} \phi_{x,A}) \\
&= \hat{h} \circ (\tilde{\prod_{x \in C}} \phi_{x,A}) \\
&= \hat{h} \circ \theta_A
\end{aligned}$$

and $\theta_A$ is the desired $U_A$ translation. □

### 5.6 The Identity Functor

These results suggest that any functor that operates on 'constructions' in a pointwise fashion is probably uniform on maps. But there is one glaring exception.

**Proposition 35.** *The identity functor* Id : $\mathbf{Set}_U \to \mathbf{Set}_U$ *is not uniform on maps.*

*Proof.* Suppose Id : $\mathbf{Set}_U \to \mathbf{Set}_U$ is uniform on maps. Then if $A \subseteq U$ then there is a mapping $\phi_A : A \to U_A$ such that $h = \hat{h} \circ \phi_A$ for all $h : A \to U$.

Let $A = \{1\}$ and define $h_1$, $h_2 : \{1\} \to U$ by $h_1(1) = 1$ and $h_2(1) = \langle 1; 1\rangle$. Then $1 = \hat{h}_1(\phi_A(1))$; by the definition of substitution, this implies $\phi_A(1) = 1$. Also $\langle 1; 1\rangle = \hat{h}_2(\phi_A(1))$; by the definition of substitution, this implies $\phi_A(1) = \langle a; b\rangle$ for some $a$, $b \in A \tilde{+} U_A$. But then $1 = \langle a; b\rangle$, which is absurd. □

An alternative proof uses the Special Final Coalgebra Theorem. If Id is uniform on maps then $J_{\mathrm{Id}}$ is a final Id-coalgebra. But a final Id-coalgebra must be a singleton set, while $J_{\mathrm{Id}} = U$ and $U$ contains 0 and 1 as elements.

This circumstance is awkward. The natural way of constructing suitable functors is to combine constant and identity functors by products, sums, etc. Since the identity functor is not uniform on maps, this approach fails. Various similar functors are uniform on maps, such as $- \tilde{\times} \mathrm{K}_{\{0\}}$ and $- \tilde{\times} -$; both have the singleton set $\{0\}$ as their greatest fixedpoint. Assuming AFA does not help; the identity functor is not uniform on maps in Aczel's system either.

## 6 Conclusions

In semantics it is not customary to worry about the construction of a particular object provided it has the desired abstract properties. From this point of view, the general theorems of Aczel and Mendler [3] and Barr [4] yield final coalgebras for a great many functors.

But there is an undoubted interest in Aczel's weaker final coalgebra theorem, proved using the Anti-Foundation Axiom (AFA) [2]. Its appeal is its concreteness. The set of streams over $A$ is simply the greatest fixedpoint of the functor $A \times -$, which is also that functor's final coalgebra. Its elements are easily visualised objects of the form $\langle a_0, a_1, a_2, \ldots\rangle$.

The original motivation for my work was to treat streams and other infinite data structures. I wished to use the standard ZF axiom system as it was automated using Isabelle. Thomas Forster suggested that Quine's treatment of ordered pairs might help. Generalizing this treatment led to the new definition of functions (and thus infinite streams), in order to compare the approach with AFA. This part of the work closely follows Aczel, and Rutten and Turi [12], from the Substitution Lemma onwards.

My Special Final Coalgebra Theorem is less general than Aczel's, especially as regards concurrency. Here is a typical example. Let $\mathcal{P}_f$ be the finite powerset operator, which returns the set of all finite subsets of its argument. Consider the set $P$ of processes defined as the final coalgebra of $\mathcal{P}_f(A \times -)$. With AFA the final coalgebra is the greatest solution of $P = \mathcal{P}_f(A \times P)$, and if $p \in P$ then

$$p = \{\langle a_1, p_1\rangle, \ldots, \langle a_n, p_n\rangle\}$$

with $n < \omega$, $a_1$, ..., $a_n \in A$ and $p_1$, ..., $p_n \in P$. My approach does not handle general set constructions, only variant tuples and functions; I do not know how to model $\mathcal{P}_f$ respecting set equalities such as $\{x, y\} = \{y, x\} = \{x, y, x\}$.

My approach works best in its original application, infinite data structures. We can model the main constructions in $U^\omega$. Since $U^\omega \subseteq V_{\omega+1}$, each infinite data structure is a subset of $V_\omega$ and thus is a set of hereditarily finite sets.[3] Section 2.1 discussed infinite streams. The set $S$ of streams over $A$ is the greatest solution of $S = A \mathbin{\tilde{\times}} S$, and is the final coalgebra of the functor $A \mathbin{\tilde{\times}} -$.

Thus we have an account of non-well-founded phenomena that is concrete enough to be understood directly. One can argue about the constructive validity of the cumulative hierarchy, but $V_\omega$ is uncontroversial even from an intuitionistic viewpoint. In contrast the general final coalgebra theorems [3, 4], with their quotient-of-sum constructions, are anything but concrete.

Aczel has shown that by adopting AFA we can obtain final coalgebras as greatest fixedpoints, dualising a standard result about initial algebras. My approach is another way of doing the same thing, though for fewer functors. Whether or not one choose to adopt AFA hinges on a number of issues: philosophical, theoretical, practical. Variant tuples and functions are a simple alternative.

[3] An *hereditarily finite set* is one built in finitely many stages from the empty set.

# A Proof of Prop. 21

Abbreviate the functor $\mathcal{Q}(X \tilde{+} -)$ as $\mathcal{Q}_X$. This simplifies the statement of the Proposition:

**Proposition.** *Let* $U_X = \bigcup_{Z \subseteq V_{\mu+1}} Z \subseteq \mathcal{Q}_X(Z)$*. Then*

*(a)* $U_X = \mathcal{Q}_X(U_X)$*.*
*(b) If* $Z \subseteq \mathcal{Q}_X(Z)$ *then* $Z \subseteq U_X$*.*
*(c)* $U_X$ *is the final* $\mathcal{Q}_X$*-coalgebra.*

*Proof of (a).* Apply the Knaster-Tarski theorem. Clearly $\mathcal{Q}_X$ is monotone; by Lemma 6, $\mathcal{Q}_X(V_{\mu+1}) \subseteq V_{\mu+1}$. □

Before continuing, we need some more elementary facts about the cumulative hierarchy.

**Lemma 36.** *If* $\alpha$ *is an ordinal then*

$$\begin{aligned} A \times B \subseteq V_\alpha & \textit{ implies } A, B \subseteq V_\alpha \\ A + B \subseteq V_\alpha & \textit{ implies } A, B \subseteq V_\alpha \\ A \tilde{\times} B \subseteq V_{\alpha+1} & \textit{ implies } A, B \subseteq V_{\alpha+1} \\ A \tilde{+} B \subseteq V_{\alpha+1} & \textit{ implies } A, B \subseteq V_{\alpha+1} \end{aligned}$$

*Proof.* The first part holds by Lemma 10 and the transitivity of $V_\alpha$. The other parts hold by similar tedious reasoning from the definitions. □

We now tackle the next part of Prop. 21.

*Proof of (b).* Use the definition of $U_X$ after first establishing $Z \subseteq V_{\mu+1}$. By the previous Lemma, it suffices to prove $X \tilde{+} Z \subseteq V_{\mu+1}$. By Lemma 7, it suffices to prove

$$\forall_{w \in X \tilde{+} Z}\, w \cap V_\alpha \subseteq V_\mu$$

for all $\alpha$. Proceed by transfinite induction on the ordinal $\alpha$.

Let $w \in X \tilde{+} Z$. There are two cases. The first case is $w = \langle 0; x \rangle = 0 + x$ for $x \in X$. Since $X \subseteq V_\mu$, clearly $0 + x \subseteq V_\mu$. The second case is $w = \langle 1; z \rangle = 1 + z$ for $z \in Z$. We must show $1 + z \subseteq V_\mu$; by Lemma 4 it suffices to show $z \subseteq V_\mu$.

Since $z \in Z$, we have $z \in \mathcal{Q}_X(Z) = \{1\} \cup (I \tilde{\to} X \tilde{+} Z)$. The case $z = 1$ is trivial. So we may assume $z = \tilde{\lambda}_{i \in I}\, w_i$, with $w_i \in X \tilde{+} Z$ for all $i \in I$. In this case we have

$$\begin{aligned} (\tilde{\lambda}_{i \in I}\, w_i) \cap V_\alpha &\subseteq \bigcup_{\beta < \alpha} \tilde{\lambda}_{i \in I}\, (w_i \cap V_\beta) \\ &\subseteq \bigcup_{\beta < \alpha} \tilde{\lambda}_{i \in I}\, V_\mu \\ &\subseteq V_\mu \end{aligned}$$

by Lemma 11, the induction hypothesis for $w_i$ and Lemma 5. Since $w \cap V_\alpha \subseteq V_\mu$ for all $\alpha$ we have $w \subseteq V_\mu$ for all $w \in X \tilde{+} Z$. This establishes $X \tilde{+} Z \subseteq V_{\mu+1}$ as required. □

The final part of Prop. 21 is that $U_X$ is the final $\mathcal{Q}_X$-coalgebra. This requires showing that for every map $f : A \to \mathcal{Q}_X(A)$ there exists a unique map $\pi : A \to U_X$ such that $\pi = \mathcal{Q}_X(\pi) \circ f$:

$$\begin{array}{ccc} A & \xrightarrow{\ \pi\ } & U_X \\ {\scriptstyle f}\downarrow & & \| \\ \mathcal{Q}_X(A) & \xrightarrow[\ \mathcal{Q}_X(\pi)\ ]{} & \mathcal{Q}_X(U_X) \end{array}$$

Let the set $A$ and the map $f : A \to \mathcal{Q}_X(A)$ be fixed, and consider each property separately. Note that the functor $\mathcal{Q}_X$ operates on maps as follows:

$$\mathcal{Q}_X(\pi)(\tilde{\lambda}_{i\in I}\, a_i) = \tilde{\lambda}_{i\in I}\, (\mathrm{id}_X \mathbin{\tilde{+}} \pi)(a_i)$$

**Lemma 37.** *There exists $\pi : A \to U_X$ such that $\pi(a) = \mathcal{Q}_X(\pi)(f(a))$ for all $a \in A$.*

*Proof.* The function $\pi$ is defined by $\pi(a) \equiv \bigcup_{n<\omega} \pi_n(a)$, where $\{\pi_n\}_{n<\omega}$ is as follows:

$$\begin{aligned} \pi_0(a) &\equiv 0 \\ \pi_{n+1}(a) &\equiv \mathcal{Q}_X(\pi_n)(f(a)) \end{aligned}$$

Suppose $a \in A$ and prove $\pi(a) = \mathcal{Q}(\pi)(f(a))$ by cases. If $f(a) = 1$ then the equation reduces to $1 = 1$. The other possibility is $f(a) = \tilde{\lambda}_{i\in I}\, b_i$ where $b_i \in X \mathbin{\tilde{+}} A$ if $i \in I$. Continuity reasoning, using the previous lemma, establishes the equation:

$$\begin{aligned} \pi(a) &= \bigcup_{n<\omega} \pi_n(a) \\ &= \bigcup_{n<\omega} \pi_{n+1}(a) \\ &= \bigcup_{n<\omega} \mathcal{Q}_X(\pi_n)(\tilde{\lambda}_{i\in I}\, b_i) \\ &= \bigcup_{n<\omega} \tilde{\lambda}_{i\in I}\, (\mathrm{id}_X \mathbin{\tilde{+}} \pi_n)(b_i) \\ &= \tilde{\lambda}_{i\in I} \bigcup_{n<\omega} (\mathrm{id}_X \mathbin{\tilde{+}} \pi_n)(b_i) \\ &= \tilde{\lambda}_{i\in I}\, (\mathrm{id}_X \mathbin{\tilde{+}} \pi)(b_i) \\ &= \mathcal{Q}_X(\pi)(\tilde{\lambda}_{i\in I}\, b_i) \\ &= \mathcal{Q}_X(\pi)(f(a)) \end{aligned}$$

Note that $\bigcup_{n<\omega}(\mathrm{id}_X \tilde{+} \pi_n)(b_i) = (\mathrm{id}_X \tilde{+} \pi)(b_i)$ above holds by case analysis on $b_i \in X \tilde{+} A$ and continuity of the variant injections.

To show $\pi : A \to U_X$, use coinduction. Let $Z = \{\pi(a) \mid a \in A\}$ and prove $Z \subseteq \mathcal{Q}_X(Z)$. If $z \in Z$ then $z = \pi(a)$ for some $a \in A$. There are two cases, as usual. If $f(a) = 1$ then $z = 1 \in \mathcal{Q}_X(Z)$. If $f(a) = \tilde{\lambda}_{i\in I}\, b_i$ (where $b_i \in X \tilde{+} A$) then

$$z = \pi(a) = \mathcal{Q}_X(\pi)(\tilde{\lambda}_{i\in I}\, b_i) = \tilde{\lambda}_{i\in I}\,(\mathrm{id}_X \tilde{+} \pi)(b_i) \in \mathcal{Q}(X \tilde{+} Z) = \mathcal{Q}_X(Z).$$

Since $U_X$ is the greatest post-fixedpoint of $\mathcal{Q}_X$, this establishes $Z \subseteq U_X$. And since $Z$ is the range of $\pi$, this establishes $\pi : A \to U_X$. □

**Lemma 38.** *If $\pi = \mathcal{Q}_X(\pi) \circ f$ and $\pi' = \mathcal{Q}_X(\pi') \circ f$ then $\pi = \pi'$.*

*Proof.* Using Lemma 7, let us use transfinite induction on the ordinal $\xi$ to prove

$$\forall_{a\in A}\, \pi(a) \cap V_\xi \subseteq \pi'(a).$$

Let $a \in A$. If $f(a) = 1$ then $\pi(a) = \pi'(a) = 1$. If $f(a) = \tilde{\lambda}_{i\in I}\, b_i$ (where $b_i \in X \tilde{+} A$) then

$$\begin{aligned}
\pi(a) \cap V_\xi &= (\tilde{\lambda}_{i\in I}\,(\mathrm{id}_X \tilde{+} \pi)(b_i)) \cap V_\xi \\
&\subseteq \bigcup_{\eta<\xi} \tilde{\lambda}_{i\in I}\,((\mathrm{id}_X \tilde{+} \pi)(b_i) \cap V_\eta) \\
&\subseteq \tilde{\lambda}_{i\in I}\,(\mathrm{id}_X \tilde{+} \pi')(b_i) \\
&= \pi'(a)
\end{aligned}$$

using the hypothesis, Lemma 11 and monotonicity of $\tilde{\lambda}$. The reasoning also uses

$$(\mathrm{id}_X \tilde{+} \pi)(b_i) \cap V_\eta \subseteq (\mathrm{id}_X \tilde{+} \pi')(b_i),$$

which holds by case analysis on $b_i \in X \tilde{+} A$, a further application of Lemma 11, and the induction hypothesis for $\eta'$ where $\eta' < \eta < \xi$.

Since $\pi(a) \cap V_\xi \subseteq \pi'(a)$ for every ordinal $\xi$, we have $\pi(a) \subseteq \pi'(a)$. By symmetry we have $\pi'(a) \subseteq \pi(a)$ and therefore $\pi(a) = \pi'(a)$ for all $a \in A$. □

**Theorem 39.** *$U_X$ is a final $\mathcal{Q}_X$-coalgebra.*

*Proof.* Immediate by the previous two lemmas. □